\documentclass[aps,prl,twocolumn,showpacs]{revtex4-1}
\usepackage{graphicx,color}

\begin{document}

\title{Effective Rheology of Immiscible Two-Phase Flow in Porous Media}

\author{Santanu Sinha}
\email{Santanu.Sinha@ntnu.no}
\author{Alex Hansen}
\email{Alex.Hansen@ntnu.no}
\affiliation{Department of Physics, Norwegian University of Science and
  Technology, N-7491 Trondheim, Norway}

\date{\today}

\begin{abstract}
We demonstrate through numerical simulations and a mean-field
calculation that immiscible two-phase flow in a porous medium behaves
effectively as a Bingham viscoplastic fluid. This leads to a
generalized Darcy equation where the volumetric flow rate depends
quadratically on an excess pressure difference in the range of flow
rates where the capillary forces compete with the viscous forces. At
higher rates, the flow is Newtonian.
\end{abstract}

\maketitle

The simultaneous flow of immiscible fluids in porous media
\cite{b88,d92,s95} lies at the heart of a wide range of important
applications ranging from oil recovery to ground water management.
Within the statistical physics community, there has been considerable
interest in this problem since the eighties when the fractal structure
of fluid invasion was discovered and explored \cite{ltz88,f88}. Such
invasion phenomena are transient and may be characterized by the
macroscopic flow parameters, such as injected pore volumes of invading
fluids change on the same time scale as those associated with the
internal flow. Much less attention, however, has been offered to {\it
  steady state flow\/} which occurs when macroscopic parameters change
slowly compared to those associated with the internal flow
\cite{gr93,ap95,ap99,kah02,kh02,kh06,rh06}.

The steady state has recently been studied experimentally by
Tallakstad {\it et al.\/} \cite{tkrlmtf09,tlkrfm09} in a
two-dimensional Hele-Shaw cell filled with glass beads. There are 15
evenly spaced tubes at one edge of the cell. Air and glycerol are
injected at equal rates through each alternate inlet. Fluids leave at
the opposing edge, which is kept open. The other two edges of the
cell, perpendicular to the direction of overall flow, are closed. As
the two immiscible fluids move along the Hele-Shaw cell, they form
interpenetrating clusters separated by interfaces. This constitutes
the steady state.

The steady state is characterized by a number of macroscopic
parameters: Capillary number ${\rm Ca}$, viscosity ratio $M$, total
volumetric flow rate $Q$, non-wetting fractional flow rate $F_{nw}$,
non-wetting saturation $S_{nw}$ and pressure $P$.  The capillary
number is the ratio between the typical viscous pressure drop across
the pores and the typical capillary force due to the interface between
the fluids.  Tallakstad {\it et al.\/} observed that the average
pressure gradient $\Delta P$ throughout the system scales as a power
law with the capillary number ${\rm Ca}$ as,
\begin{equation}
\label{beta}
\Delta P \sim {\rm Ca}^{\beta},
\end{equation}
where $\beta = 0.54\pm0.08$.  

More recently, Rassi {\it et al.\/} \cite{rcs11} have measured the
exponent $\beta$ which varies in the range of $0.3$ to $0.45$
depending on the saturation, in steady-state two-phase flow of water
and air in a three-dimensional porous medium constructed from glass
beads.

These observations have profound implications on the description of
multiphase flow in porous media. An important application would be in
the reservoir simulators, used in the exploitation of oil reservoirs.
They are based on effective transport equations for the fluids where a
linear relation between the pressure gradients and flow rates
\cite{b88} is assumed so far.

We will in this Letter demonstrate that, in the regime where capillary
forces are comparable to the viscous forces (low ${\rm Ca}$), the
capillary effects at the interfaces between the immiscible fluids
effectively creates a yield threshold, making the fluids reminiscent
of a Bingham viscoplastic fluid \cite{ld84,rh87} in the porous medium
--- i.e.\ a fluid possessing a yield stress and a constant effective
viscosity. This introduces a overall threshold pressure $\Delta P_c$
in the system due to the random distribution of capillary pressure
thresholds. We therefore propose, and will show in the following via
numerical simulations and  a mean field calculation, that the
steady-state two-phase flow in porous media in this flow regime is 
governed by a generalized Darcy equation 
\begin{widetext}
\begin{equation}
\label{effdarcy}
Q = -\ C\ \frac{A}{L}\ \frac{K(S_{nw})}{\mu_{\textrm{eff}}(S_{nw})}
{\rm sgn}\left(\Delta P\right)\
\left\{  \begin{array}{cl} 
\left(|\Delta P|-\Delta P_c(S_{nw})\right)^2 
  & \mbox{if $|\Delta P| >   \Delta P_c$}\;,\\
0 & \mbox{if $|\Delta P| \le \Delta P_c$}\;,\\
\end{array}
\right.
\end{equation}
\end{widetext}
where $\rm sgn$ is the sign function. Here $A$ is the cross section of
the representative elementary volume, $L$ is its length, $C$ a
constant with units of inverse pressure, $K(S_{nw})$ the effective
permeability which depends on the saturation $S_{nw}$ and
$\mu_{\textrm{eff}}(S_{nw})$ is the saturation-weighted viscosity
given by $S_{nw} \mu_{nw}+(1-S_{nw})\mu_w$, where $\mu_w$ and $\mu_{nw}$ are
the viscosities of wetting and non-wetting fluids respectively. 

As a result, the correct scaling relation in general between $\Delta P$ and
${\rm Ca}$ is not Eq.\ (\ref{beta}), but
\begin{equation}
\label{sclPc}
(|\Delta P|-\Delta P_c)\sim {\rm Ca}^\beta\;,
\end{equation}
with $\beta=1/2$ for low ${\rm Ca}$. The relation (\ref{beta})
is just a special case of this relation where $\Delta P_c \approx 0$.

There is another flow regime for high ${\rm Ca}$ where the flow is
linear with the excess pressure drop which corresponds to
$\beta=1$. This regime is therefore characterized by the
standard Darcy equation $Q=-(A/L)(K/\mu_{\rm eff}(S_{nw})) \Delta
P$. Here $|\Delta P| \gg \Delta P_c$ and the threshold pressure is not
relevant.

The physical existence of the global threshold pressure $\Delta P_c$
and the quadratic and linear dependence of $Q$ on $(|\Delta P| -
\Delta P_c)$ in the two regimes for the system of immiscible fluids
can be understood very intuitively following the argument of Roux and
Herrmann \cite {rh87} for networks with link
conductances having characteristics like a Bingham fluid: The
essential ingredient is a threshold pressure for each link,
distributed according to some probability density. The sum over all
the thresholds over a continuous flow path throughout the entire
system gives the total threshold pressure along that path and $\Delta
P_c$ corresponds to the minimum sum among all such possible
paths. Now, if we raise the pressure difference across the network by
a value $dP$, a number $d{\cal N}$ of tubes will cross their flow
thresholds. With a reasonably smooth distribution of thresholds, we
will have that $d{\cal N}\propto dP$. The conductance of the network
$\Sigma$ will then change by an amount $d\Sigma$, and $d\Sigma\propto
d{\cal N}\propto dP$. An integration over this equation leads to
Eq.\ (\ref{effdarcy}) with the appearance of $\Delta P_c$. Moreover,
at a very high value of $P$, when all the links have crossed their
individual flow thresholds, each link behaves linearly with excess
pressure drop, making $\Sigma$ constant with $P$ and effectively the
overall flow rate becomes linear with the excess pressure drop in the
high ${\rm Ca}$ regime.

In the following, we will first present our numerical
simulations. Afterwards, we will derive the effective Darcy equation
(\ref{effdarcy}), adapting the mean-field calculation of the
conductivity of heterogeneous conductors by Kirkpatrick \cite{k73}. 

In our numerical studies, the porous medium is modeled by a
two-dimensional network of tubes, forming a square lattice tilted by
$45^\circ$ with respect to the imposed pressure gradient. Two
immiscible fluids, one is more wetting than the other with respect to
the pore walls, flow inside the tubes. Disorder is introduced in the
system by choosing the radius $r$ of each tube randomly from a uniform
distribution of random numbers in the range $[0.1l, 0.4l]$, where $l$
is the length of the tubes. In order to incorporate the shape of the
pores in between spherical particles (beads) that introduces the
capillary effect in the system, each tube is considered hour-glass
shaped so that the capillary pressure $p_c$ at a meniscus at position
$x$ is proportional to $[2\gamma/r][1-\cos(2\pi x/l)]$ where $\gamma$
is the surface tension \cite{d92,amhb98}.

The flow is driven by setting up an external global pressure drop. The
local flow rate $q$ in a tube with a pressure difference $\Delta p$
between the two ends of that tube follows the Washburn equation of
capillary flow \cite{d92,amhb98}
\begin{equation}
\displaystyle
q = -\frac{a k}{\mu_{\textrm{eff}}(s_{nw}) l}\left(\Delta p - \sum p_c\right)
=-\sigma_0 \left(\Delta p-\Sigma p_c\right)\;,
\label{wb}
\end{equation}
where $k=r^2/8$ is the permeability for cylindrical tubes. The conical
shape of the tubes leads only to an overall geometrical factor. Here
$a$ is the cross-sectional area of the tube and
$\mu_{\textrm{eff}}(s_{nw})$ is the volume average of the viscosities
of the two phases present inside the tube. Hence, it is a function of
the saturation $s_{nw}$ in the tube. The sum over $p_c$ runs over all
menisci within the tube. $\sigma_0$ is thus the tube conductivity.
The set consisting of one equation (\ref{wb}) per tube, together with
the Kirchhoff equations balancing the in and out flow at each node are
then solved using Cholesky factorization combined with a conjugate
gradient solver. The system is then integrated in time using an
explicit Euler scheme. Inside a tube all menisci move with a speed
determined by $q$. When a meniscus reaches the end of a tube, new
menisci are formed in the neighboring tubes. Further details of the
model and how the menisci are moved can be found in
\cite{amhb98,kah02}.


The steady-state condition in the simulation is achieved in two
ways. The conventional way is to use bi-periodic boundary conditions
(BP), by connecting the inlet and outlet rows so that the network
takes a toroidal topology \cite{kh02}. It is then initialized by
filling with two fluids randomly or sequentially so that the network
attains the desired saturation $S_{nw}$. As the system is closed in
this boundary condition, the saturation of the network $S_{nw}$ is an
independent parameter which remains constant throughout the simulation
along with the total flow rate $Q$ whereas the fractional flow
$F_{nw}$ fluctuates over time.

It is not possible to implement bi-periodic boundary condition in the
experiments by Tallakstad {\it et al.\/} \cite{tkrlmtf09,tlkrfm09}, where
two fluids are injected at one edge of the system through a series of
alternate inlets and the opposite edge is kept open. Flow rates of the
two fluids may be controlled independently there. The control
parameters in this case are the total flow rate $Q$ and the fractional
flow $F_{nw}$ whereas the saturation $S_{nw}$ fluctuates. Therefore,
in order to have a close emulation of the experimental system, we also
implement open boundary conditions (OB) in our simulation here,
controlling the individual flow rates of inlet links. The simulation
starts with injecting the two fluids with constant flow rates in a
system completely saturated with the wetting fluid. Both drainage and
imbibition therefore takes place at the pore level creating new
menisci. Away from the inlets, the fluids mix and a steady state is
attained as in the experiment.

\begin{figure}
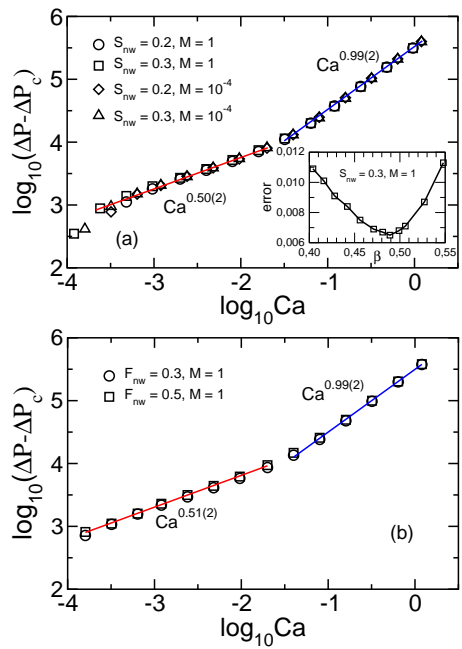

\includegraphics[scale=0.31,clip]{fig1a.eps}
\includegraphics[scale=0.31,clip]{fig1b.eps}
\caption{\label{fig2} Scaling of excess steady-state pressure drop
  ($|\Delta P| - \Delta P_c$) (in Pa) with cillary number ${\rm Ca}$
  for (a) BP and (b) OB boundary conditions. The threshold pressure
  $\Delta P_c$ have determined from minimizing the least square fit
  errors. This minimization is shown as a function of $\beta$ in the
  inset of (a).}
\end{figure}

Simulations are performed with constant flow rate $Q$ which sets the
capillary number ${\rm Ca}$ given by ${\rm Ca} = \mu_{\textrm{eff}}
Q/(\gamma A)$. A range of ${\rm Ca}$ from $10^{-4}$ to $1$ is
considered for a network of $64\times 64$ links and an average over
$10$ different samples is taken for each simulation. We report results
for $M = 1$ and $10^{-4}$ with $S_{nw} = 0.2$ and $0.3$ for PB, and $M
= 1$ with $F_{nw} = 0.3$ and $0.5$ for OB. The steady state is
identified from the total pressure drop $\Delta P$, which starts to
fluctuate over an average value as steady state is reached. By
increasing and then lowering the total flux $Q$ we also verify that it
returns to the same steady state. In order to verify
Eq.\ (\ref{sclPc}), we now need to calculate $\Delta P_c$. Roux
and Herrmann \cite {rh87}, for their Bingham system, determined
$\Delta P_c$ by decreasing the external current from a large value and
identifying the current paths by a search algorithm. This procedure is
not feasible here as the flow patterns and menisci positions change
with global flow rate and time due to fluid instabilities. Moreover,
it is also not possible to follow this in experiments as it would
necessitate the knowledge of flow rates at every single pore. We
therefore measure $\Delta P_c$ with a minimization procedure. A series
of trial $\Delta P_c$ values are considered, for which the slope
($\beta$) and the least square fit errors, when fitted to
Eq.\ (\ref{sclPc}), are calculated. $\Delta P_c$ is then identified
corresponding to the minimum value of the error or the best fit. This
is shown in the inset of Fig.\ \ref{fig2}(a) for $M=1$ and
$S_{nw}=0.3$. In Table \ref{pcval}, the absolute values of all $\Delta
P_c$s, identified by the same procedure, are listed. In
Fig.\ \ref{fig2}, $(|\Delta P| - \Delta P_c)$ is then plotted with
${\rm Ca}$ for (a)BP and (b)OB according to
Eq.\ (\ref{sclPc}). Interestingly, for different saturations
($S_{nw}$), fractional flows ($F_{nw}$) and boundary conditions, the
minimum error corresponds to different values of $\Delta P_c$ but the
same value for $\beta = 0.5$ within error bar for low ${\rm Ca}$
regime. More surprisingly, $\Delta P_c$ is found independent to the
viscosity ratio which is a strong support towards its intuitive
physical description stated before. As $\Delta P_c$ is related to the
sum of capillary thresholds ($p_c$) over connecting paths and $p_c$
does not depend on the viscosities, $\Delta P_c$ also should not and
that is what we found here. A sharp crossover is also seen for high
${\rm Ca}$ regime with $\beta = 1$ where the flow is linear
characterized by standard Darcy equation.

\begin{table}
\centerline{\hfill (a) Biperiodic (BP) \hfill \hfill (b) Open (OP) \hfill}
\hfill
\begin{minipage}[t]{.2\textwidth}
\vspace{0pt}
\centering
\begin{tabular}{c|c|c}
$S_{nw}$ & $M$ & $P_c$ (KPa) \\
\hline
$0.2$ & $1$ & $3.45 \pm 0.05$ \\
$0.3$ & $1$ & $5.10 \pm 0.05$ \\
$0.2$ & $10^{-4}$ & $3.45 \pm 0.05$ \\
$0.3$ & $10^{-4}$ & $5.10 \pm 0.05$ \\
\end{tabular}
\end{minipage}%
\hfill
\begin{minipage}[t]{.2\textwidth}
\vspace{0pt}\raggedright
\begin{tabular}{c|c|c}
$F_{nw}$ & $M$ & $P_c$ (KPa) \\
\hline
$0.3$ & $1$ & $6.55 \pm 0.05$ \\
$0.5$ & $1$ & $6.15 \pm 0.05$ \\
\end{tabular}
\end{minipage}
\caption{\label{pcval}Values of threshold pressures ($\Delta P_c$)
  measured by minimizing the least square fit errors for different
  parameters and boundary conditions.}
\end{table}

In the experiments by Rassi {\it et al.\/} \cite{rcs11}, the exponent
$\beta$, when measured according to Eq.\ (\ref{beta}) is found to vary
from $0.30$ to $0.45$ depending on the saturation. A preliminary
numerical study \cite{gh11} assuming Eq.\ (\ref{beta}) using a very
similar model with only BP, also reports similar dependency of $\beta$
on saturation with a visible curvature in the scaling plots. These
clearly indicate a non-zero $\Delta P_c$ that has been ignored hence
resulting in a wandering value of $\beta$. Interestingly, the same
experimental data by Rassi {\it et al.\/} \cite{rcs11} are found
consistent with $\beta=1/2$ when reanalyzed using Eq.\ (\ref{sclPc})
\cite{c12}. In the experiment by Tallakstad {\it et al.\/}
\cite{tkrlmtf09,tlkrfm09}, $\beta$ is found as $0.54\pm0.08$ with
scaling relation (\ref{beta}) which is due to the fact that one of the
fluids was percolating in their system \cite{m12}, making
$\Delta P_c=0$. 

In support of our numerical results, we now derive the generalized
Darcy equation, Eq.\ (\ref{effdarcy}) in a mean field
approximation. Eq.\ (\ref{wb}) describes instantaneous the flow in a
single tube.  Time averaging this equation under steady state
conditions leads to an effective flow equation for the single tube
\cite{shbk12}
\begin{equation}
\label{kirk0}
q = - \sigma_0\ {\rm sgn}(\Delta p)
\left\{  \begin{array}{cl} 
\sqrt{\Delta p^2-\Delta p_c^2}
  & \mbox{if $|\Delta p| >   \Delta p_c$}\;,\\
0 & \mbox{if $|\Delta p| \le \Delta p_c$}\;.\\
\end{array}
\right.
\end{equation}
where $\Delta p_c$ is an effective flow threshold that depends on the
shape of the tube. The effective viscosity that enters into $\sigma_0$
is the saturation-weighted sum of the viscosities of each liquid,
where the saturation is time averaged over the tube.

The square root singularity near $\Delta p_c$ is caused by a
saddle-node bifurcation and, hence, is a universal feature of the
system \cite{s94}.

The effective conductivity is thus
\begin{equation}
\label{effsigma}
\sigma(\Delta p)=-\frac{dq}{d(\Delta p)}=\sigma_0\ 
\left\{  \begin{array}{cl} 
\frac{|\Delta p|}{\sqrt{\Delta p^2-\Delta p_c^2}}
  & \mbox{if $|\Delta p| >   \Delta p_c$}\;,\\
0 & \mbox{if $|\Delta p| \le \Delta p_c$}\;.\\
\end{array}
\right.
\end{equation}  

In the following, we will derive Eq.\ (\ref{effdarcy}) based on the
mean-field theory originally developed for calculating the
conductivity of percolating systems by Kirkpatrick \cite{k73}. Our
starting point is a regular lattice with coordination number $z$.
Each tube in the lattice has a non-linear conductivity $\sigma(\Delta
p)$, Eq.\ (\ref{effsigma}). The flow thresholds of the tubes are drawn
from a spatially uncorrelated probability distribution $\pi(\Delta
p_c)$.

The mean-field calculation proceeds by focusing on one tube inside the
network.  We then replace the rest of the network by an {\it equivalent 
homogeneous network\/} where all tubes have the same conductivity 
$m(\Delta p)$ so
that the tube we have singled out experiences the same average effect
from the homogeneous network as from the original network. We then
average over this last conductance and determine $m(\Delta p)$ 
in a self-consistent way.  

There is one caveat. In Kirkpatrick's original calculation, the conductances
were assumed to be linear, whereas in our case, the conductances are
highly non-linear, see Eq.\ (\ref{effsigma}).  However, the end result of the
calculation,    
\begin{equation}
\label{kirk4}
\left\langle\frac{m(\Delta p)-\sigma(\Delta p)}{\left(\frac{z-2}{2}\right)
m(\Delta p)+\sigma(\Delta p)}\right\rangle=0\;,
\end{equation}
is the same.  This equation provides a self-consistent expression for the 
equivalent conductivity $m(\Delta p)$.

In terms of the distribution of flow thresholds $\pi(\Delta p_c)$, 
Eq.\ (\ref{kirk4}) becomes
\begin{equation}
\label{kirk5}
\int_0^\infty dp\ \pi(p)\ \frac{m(\Delta p)-\sigma(\Delta p)}
{\left(\frac{z-2}{2}\right)m(\Delta p)+\sigma(\Delta p)}=0\;.
\end{equation}
We now combine this expression with Eq.\ (\ref{effsigma}), finding
\begin{widetext}
\begin{equation}
\label{kirkb}
\int_0^{\Delta p} dp\ \pi(p)\
\frac{m(\Delta p)\sqrt{\Delta p^2-p^2}
-\sigma_0|\Delta p|}{\left(\frac{z-2}{2}\right) 
m(\Delta p) \sqrt{\Delta p^2-p^2}+\sigma_0|\Delta p|}
+ 1 - \Pi(\Delta p) = 0\;,
\end{equation}
\end{widetext}
where $\Pi(p)= \int_0^p dp'\ \pi(p')$ is cumulative probability 
to find a flow threshold less than or equal to $p$. 

By setting $m(\Delta p)=0$ in Eq.\ (\ref{kirkb}), we determine the effective 
flow threshold $\Delta \overline{p}_c$ for the effective tubes:
\begin{equation}
\label{kirkc}
\Pi(\Delta \overline{p}_c)=\frac{1}{2}\;.
\end{equation}    

If we now set $\Delta p > \Delta\overline{p}_c$ in Eq.\ (\ref{kirkb}) and
expand to lowest order in $m(\Delta p)$, we find
\begin{equation}
\label{meff}
m(\Delta p) = \frac{4\sigma_0}{8-z}\ 
\pi(\Delta\overline{p}_c)\ \frac{(\Delta p-\Delta\overline{p}_c)}
{\int_0^{\Delta\overline{p}_c} dp\
\pi(p) \sqrt{1-\left(\frac{p}{\Delta\overline{p}_c}\right)^2}}\;,
\end{equation}
where we have used that 
$\Pi(\Delta p) -1/2=\pi(\Delta\overline{p}_c)(\Delta p -\Delta\overline{p}_c)$ 
to lowest order. We integrate Eq.\ (\ref{meff}) and find
\begin{widetext}
\begin{equation}
\label{effq}
\overline{q} = -\ \frac{2\sigma_0}{8-z}\ \frac{\pi(\Delta\overline{p}_c)\
{\rm sgn}\left(\Delta p\right)}
{\int_0^{\Delta\overline{p}_c} dp\
\pi(p) \sqrt{1-\left(\frac{p}{\Delta\overline{p}_c}\right)^2}}
\left\{  \begin{array}{cl} 
\left(|\Delta p|-\Delta\overline{p}_c\right)^2 
  & \mbox{if $|\Delta p| >   \Delta\overline{p}_c$}\;,\\
0 & \mbox{if $|\Delta p| \le \Delta\overline{p}_c$}\;.\\
\end{array}
\right.
\end{equation}
\end{widetext}
If there are $N$ tubes in a cross section
of the homogeneous network orthogonal to the flow direction, then 
$Q=N\overline{q}$, 
$\Delta P = (L/l) \Delta p$ and $\Delta P_c = (L/l) \Delta\overline{p}_c$. 
Hence, the generalized Darcy equation (\ref{effdarcy}) follows.

There are corrections associated with larger exponents to
Eq.\ (\ref{meff}) originating from two sources.  The first one comes from
solving Eq.\ (\ref{kirkb}).  The prefactor of the leading correction
from this source is about 7\% of the prefactor of the dominating
term. The second source of correction terms comes from the
linearization of $\Pi(\Delta p)-1/2$ around the value
$\Delta\overline{p}_c$.  There are no reasons to assume strong
non-linear corrections in this region.

To summarize, we have demonstrated numerically and through a mean
field calculation that steady-state immiscible two-phase flow in a
porous medium behaves similar to a Bingham viscoplastic fluid. This
leads to a non-linear Darcy equation where the volumetric flow rate
depends {\it quadratically\/} on an excess pressure drop at capillary numbers
at which the capillary forces compete with the viscous forces. At
higher flow rates, the flow becomes Newtonian.

We thank D.\ Bedeaux, E.\ G.\ Flekk{\o}y, S.\ Kjelstrup, K.\ J.\ M{\aa}l{\o}y 
and L.\ Talon for useful discussions.  This work was partially supported by 
the Norwegian Research Council through grant no.\ 193298.  We thank 
NOTUR for allocation of computer time.

\end{document}